\documentclass[preprint,times,tighten]{aastex631}

\newcommand{\nc}{\newcommand}
\newcommand{\CII}{[C {\sc ii}]}
\newcommand{\HII}{H {\sc ii}}
\nc{\thCO}{$^{13}$CO}
\nc{\CeiO}{C$^{18}$O}
\nc{\cmcub}{\mbox{cm$^{-3}$}}
\nc{\cmsq}{\mbox{cm$^{-2}$}}
\nc{\Kkms}{\mbox{K~km/s}}
\nc{\kms}{\mbox{km~s$^{-1}$}}
\nc{\msun}{\ensuremath{M_\odot}}
\nc{\lsun}{\ensuremath{\mathrm{L}_\odot}}
\def\cplus{C$^{+}$}
\nc{\tkin}{\mbox{$T_{\rm kin}$}}
\newcommand{\OI}{[O {\sc i}]}
\newcommand{\thCII}{[$^{13}$C {\sc ii}]}

\submitjournal{APJ}

\shorttitle{Star Formation Triggered by expanding S111}
\shortauthors{B. Mookerjea}
\graphicspath{{./}{figures/}}

\begin{document}

\title{Star Formation Triggered by the expanding bubble S111}

\author[0000-0003-1766-6303]{Bhaswati Mookerjea}
\affiliation{Tata Institute of Fundamental Research,\\
Homi Bhabha Road,
Mumbai 400005, India}

\begin{abstract}

This paper investigates the impact of radiative and mechanical feedback
from O-type stars on their parent molecular clouds and the triggering of
formation of future generation of stars. We study the infrared bubble
S111 created  by the embedded massive stellar cluster G316.80-0.05.  A
significant fraction of gas in shells created due to the compression of
the ambient medium by expanding bubbles is photodissociated by the
stellar radiation.  The kinematics of the shells are thus best studied using
spectroscopic observations of singly ionized carbon, the most dominant
species.  We have used the velocity-resolved maps of the
$^2{\rm P}_{3/2}\rightarrow ^2{\rm P}_{1/2}$ transition of \CII\ at
158\,\micron, the
$J$=2--1 transition of \thCO\ and \CeiO, and the $J$=1--0 transition of
HCO$^+$ to study the rim of the bubble S111 that partly coincides with
the southern part of the infrared dark ridge G316.75. The \CII\ spectra
conclusively show evidence of a shell expanding with a moderate velocity
of $\sim 7$\,\kms, which amounts to  a kinetic energy that is $\sim
$0.5--40 times the thermal energy of the \HII\ region.  The pressure
causing the expansion of the \HII\ region arises mainly from the
hydrogen ionization and the dust-processed radiation.  Among the
far-infrared sources located in the compressed shells, we find the core
G316.7799-0.0942 to show broad spectral features consistent with outflow
activity and conclude that it is a site of active star formation.  Based
on the age of the \HII\ region we conclude that this expanding \HII\
region is responsible for the triggering of the current star formation
activity in the region.

\end{abstract}


\keywords{ISM: Bubbles -- infrared:~ISM -- ISM: lines and bands
--(ISM:) photon-dominated region (PDR) --ISM: individual (S111) --ISM:
individual (G316.75) -- ISM: kinematics and dynamics}

\section{Introduction} \label{sec:intro}

Massive stars play a dominant role in the injection of UV radiation into
the interstellar medium (ISM) and of mechanical energy through stellar
winds and finally through supernova explosions. The photoionization,
radiation pressure, and stellar winds from young O-type stars pump both
energy and momentum into the ambient material and disrupt the parent
molecular cloud by creating \HII\ regions and parsec-size bubbles
enclosed by shells of denser swept-up material \citep[][and references
therein]{churchwell2006}. It is envisaged that such radiative and
mechanical feedback from the massive stars regulate the star forming
environment locally as well as influence the evolution of the ISM in the
galaxies as a whole \citep{Krumholz2018}. Inventories of such
bubbles have been primarily derived from large-scale mid-infrared continuum
images observed using the Spitzer and AKARI telescopes
\citep{Deharveng2010,Hanaoka2020}. 

\begin{figure}
\centering
\leavevmode
\includegraphics[width=0.4\textwidth]{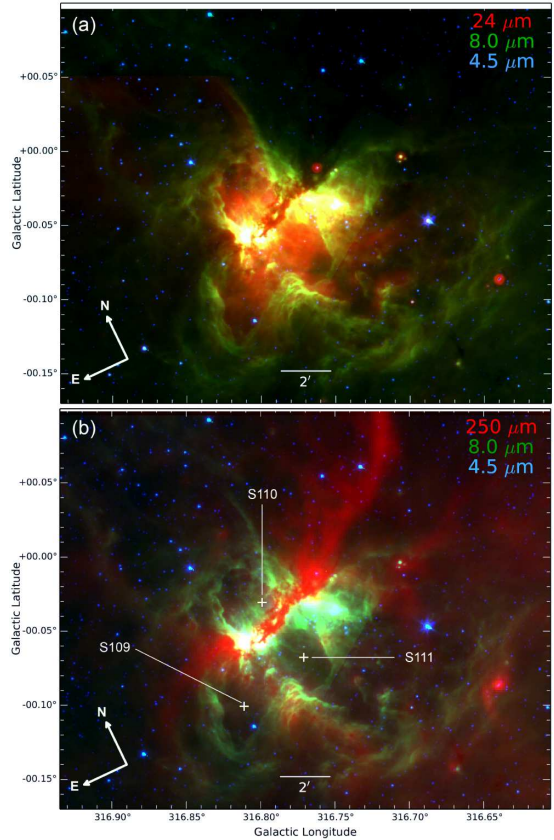}
\caption{Bipolar nebula G316.80-00.05. (a) Composite color
image of the G316.790-0.045 complex: red, green, and blue indicate the
Spitzer 24, 8.0, and 4.5\,\micron\ data, respectively 
(the 24\,\micron\ emission is saturated in the center). (b) 
Composite color image of the complex: red is for the Herschel 250\,\micron\ 
emission showing the cold filament at the waist of the nebula, 
green is for the 8\,\micron\ PAH emission, and blue is for the 4.5\,\micron\
stellar emission. The three bubbles listed in the
\citep{churchwell2006} catalog are marked. Figure reproduced from
\citet{Samal2018} with permission.
\label{fig_composite}}
\end{figure}

The triple bubble system consisting of S109, S110, and S111
\citep{churchwell2006} is created by the bipolar ionized nebula
G316.80-00.05 oriented approximately northeast to southwest as well an
older outflow extending to the east \citep{dalgleish2018}.  The parent
molecular cloud in this region is primarily in the form of a ridge that
hosts the embedded massive stellar cluster associated with the \HII\
region.  The ridge is seen in absorption in the visible, near-infrared,
and at Spitzer-GLIMPSE wavelengths \citep[Fig. B.1a in][]{Samal2018},
and is clearly detected in emission at the Herschel-SPIRE wavelengths
and in the 870\,\micron\ ATLASGAL images.  In this system the S110
bubble is the northern lobe, which extends perpendicular to the
filament, while S109 and S111 form the southern lobe
(Fig.\,\ref{fig_composite}). The complex morphology of the region is
attributed to the inhomogeneous density distribution that causes the
ionization front to be distorted and enables leakage of ionizing
photons to large distances from the embedded cluster as evidenced by
the presence of diffuse emission from the ionized gas in S110 and S111
\citep{Samal2018}.  The dust temperature distribution derived from the
Herschel Hi-GAL \citep{molinari2010} continuum observations suggests
that while the parental filament is cold ($\sim 17$\,K), going down to
14\,K, the photon dominated regions (PDRs) register temperatures well
above 20\,K \citep{Samal2018}. The column density at the position of
peak temperature position is $\sim 10^{23}$\,\cmsq, which corresponds
to a visual extinction of $\sim 100$\,mag, which explains why the
stellar cluster ionizing the region is not detected in the visible and
near-IR.

In this paper, we refer to the ridge as G316.75 and the \HII\ region as
G316.80-00.05.  The G316.75 ridge located at a distance of
2.69$\pm$0.45\,kpc \citep{Watkins2019} extends over 13.6 parsec.  The
southern part of the ridge, which is associated with an infrared dark
cloud \citep[SDC316.786-0.044][]{Peretto2009}, is more active than the
northern part \citep{Watkins2019}.  \citet{Shaver1981} were the
first to confirm the near kinematic distance of G316.75 and determined
that the source of the \HII\ region to be an O6-type star. Using
radiative transfer modeling of near- to far-infrared dust continuum
emission, \citet{vig2007} had estimated that two O-type stars of masses
45\,\msun\ and 25\,\msun\ are responsible for most of the infrared
luminosity of G316.75. This is also corroborated by high-angular-resolution radio continuum and hydrogen recombination line observations
\citep{longmore2009} which detected two 24\,GHz continuum sources at
resolutions better than 10\arcsec.  \citet{Watkins2019} have
subsequently concluded that two high-mass stars with spectral types
of O8.5--O7\,V and O6.5--O6\,V can explain the observed emission
from the ionized gas from the region.  \citet{dalgleish2018} have
studied the kinematics of the ionized gas using radio recombination line
(RRL) emission, and concluded that the strong velocity gradient they
observe could be a relic of the cloud’s initial angular momentum.
Study of the southern part of the ridge G316.75 is rendered interesting
by the fact that is a very active and young star-forming region,
harboring water, hydroxyl, and methanol masers, several O-stars, and a
compact X-ray source, and it is characterized by very dynamic gas
conditions.

Beyond the ionized gas, a significant fraction of the neutral (atomic
and molecular) material in the vicinity of such embedded clusters in
the bubbles is impacted by the far-ultraviolet (FUV:6--13\,eV)
radiation creating the so-called PDRs. In the far-infrared (FIR) the
most important cooling lines from PDRs are the fine structure lines of
\CII\ at 158\,\micron\  and \OI\ at 63 and 145\,\micron,\ and to a
lesser extent, high-$J$ CO lines, while polycyclic aromatic hydrocarbon
(PAH) emission and H$_2$ lines dominate in the near- and mid-IR. The
ionization potential of carbon being 11.26\,eV, \cplus\ is expected to
exist in ionized, atomic, and molecular gas. As a result \CII\ has a
unique position as a tracer of PDR gas owing to the ubiquity of carbon
and the moderate excitation conditions (E$_{\rm up}$=92\,K, $n_{\rm
cr}$=3000\,\cmcub) of the transition. Thus, \CII\ at 158\,\micron\ is
an ideal tool for tomographic study of the morphology and kinematics of
photoirradiated bubbles.  The availability of far-infrared observing
facilities with multi-pixel receivers has recently enabled
study of the velocity fields in the Galactic PDRs using the information
derived from mapping observations of \CII\ at 158\,\micron.

In this paper we study the morphology and kinematics of the PDR gas
in the G316.75 region that is exposed to the FUV radiation from the
embedded massive stellar cluster that also gives rise to the bipolar
\HII\ nebula G316.8-0.05. For this purpose we primarily use the velocity
information derived from observations of \CII\ emission at 158\,\micron\
in combination with $^{13}$CO(2--1) and HCO$^+$(1--0) emission that
arises primarily from the molecular material.

\section{Observations}

The paper makes use of several sets of publicly available data, which
are described below.

\subsection{GREAT/SOFIA observations}

We have retrieved observations  of the
$^2$P$_{3/2}$$\rightarrow$$^2$P$_{1/2}$ fine structure transition of ionized
carbon (\cplus) at 1900.5369\,GHz (157.74\,\micron)  of the region
around the G316.8-00.05 from the data archive of the Stratospheric
Observatory for Infrared Astronomy \citep[SOFIA;][]{young2012}.  The
observations (Id: 06\_0074; PI: J.  Jackson) were carried out using the
German REceiver for Astronomy at Terahertz frequencies
\citep[GREAT;][]{heyminck2012} on 2018 June 22.  The beam size for
\CII\ was 14\farcs1.  The \CII\ map extends over a
500\arcsec$\times$600\arcsec\ area and observations were carried out in
total power on-the-fly mode. We used the Level 4 data available on the
SOFIA archive for the region. All data were smoothed to a spectral
resolution of 0.25\,\kms\ resulting in an rms of 0.8\,K. The \OI\
63\,\micron\ fine structure transition was also mapped simultaneously as
part of the same project. However, since the mapping was optimized for
\CII, the \OI\ data were undersampled and smeared in the scan direction
and hence were not used in this work.

\subsection{SEDIGISM}

We have used the data cubes for $J$=2--1 transition of \thCO\ and
\CeiO\ that were observed as a part of the large-scale (84 deg$^2$)
spectroscopic survey of the inner Galactic disk, named Structure,
Excitation and Dynamics of the Inner Galactic Interstellar Medium
\citep[SEDIGISM;][]{schuller2021}. These data were observed with the APEX
telescope between 2013 and 2016 with an angular resolution of 30\arcsec\
and a 1$\sigma$ sensitivity less than 1.0\,K at 0.25\,\kms\ velocity
resolution.

\subsection{MALT90}

The Millimetre Astronomy Legacy Team 90 GHz (MALT90) survey
\citep{Foster2011,Foster2013,Jackson2013} mapped a sample of high-density cores
in 16 transitional lines including the high-density tracer HCO$^+$(1--0) at
89188.5234\,MHz. These data were taken with the 22\,m Mopra radio telescope by
individually observing each of the $\sim 2000$ targeted clumps in a
3\arcmin$\times$3\arcmin\ data cube. The data cubes have an angular
resolution of 38\arcsec\ and an rms noise of 0.2\,K at a spectral
resolution of $\sim 0.11$\,\kms.  Three MALT90 observations lie within the
region studied here. We have in particular used the data for HCO$^+$(1--0)
with a critical density of 2$\times 10^5$\,\cmcub\  to understand the
kinematics of the underlying dense gas in the region.

\section{Impact of Massive O-stars in G316.75 on the Ridge}

\citet{Watkins2019} have performed a detailed analysis of the radio
continuum emission from the embedded G316.75 stellar cluster that
irradiates the ridge and its vicinity. The mass of the stellar cluster
is 930$\pm$230\,\msun\ and it consists of four O-type stars one of
which has a mass larger than 48\,\msun. The ionizing stars responsible
for the two radio continuum peaks seen in the Sydney University
Molongolo Sky Survey \citep[SUMSS][]{Mauch2003} were further
constrained to be due to two high-mass stars with spectral types
O8.5--O7 and O6.5--O6. The dynamical age of a G316.75-like \HII\ region
with a radius of 6.5\,pc and a Lyman continuum photon production rate,
$N_{\rm Lyc}$ = 10$^{49.63}$\,s$^{-1}$ was estimated to be $\sim
2$\,Myr.

Earlier studies using primarily tracers of ionized and molecular material
concluded that the G316.75 ridge is mostly unaffected by ionization owing
to the large electron and H$_2$ gas densities in the immediate vicinity of
the ionizing high-mass stars \citep{Watkins2019}. These authors argued
that ionization due to OB stars happened after the formation of the
molecular ridge.  Here we concentrate on the distribution and velocities
of the FUV-irradiated gas as traced by the PDR tracer -- the \CII\ line at
158\,\micron\  -- to study the radiative and mechanical feedback of the
ionizing stars primarily on the S111 bubble.

\section{Results}

\begin{figure*}
\plotone{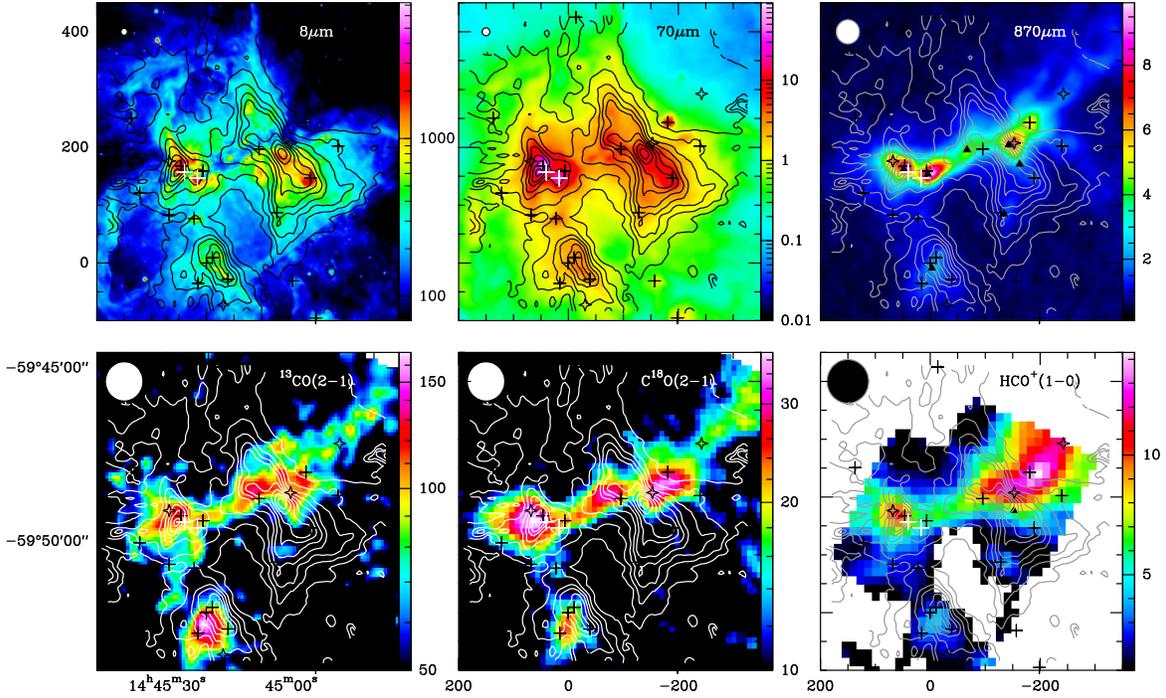}
\caption{Comparison of \CII\ emission at 158\,\micron\ emission
integrated between -65 and -20\,\kms\ (contours) with emission due to
other tracers of dust, PAHs, and molecular gas (in color and names
marked in each panel). The \CII\ contours are drawn from 90 to
540\,K\,\kms\ in steps of 50\,\kms. The color scale corresponding to
each panel is included to the right of it. The units are MJy\,sr$^{-1}$
for the 8\,\micron\ image, Jy\,pixel$^{-1}$ for 3\arcsec\ pixels for
the PACS 70\,\micron\ image, and Jy\,beam$^{-1}$ with a beam size of
21\,\arcsec\ for the ATLASGAL 870\,\micron.  The color scales of
spectroscopic images are all in units of K\,\kms. The velocity
intervals of integration of \thCO(2--1), \CeiO(2--1) and HCO$^+$(1--0)
are -60 to -10\,\kms, -47 to -20\,\kms\ and -48 to -26\,\kms,
respectively. For each of the tracers shown as color images the beams
are shown in the top left corner of each panel. The radio continuum
sources are marked with "+" (white), the ATLASGAL clumps
\citep{Csengeri2014} are marked with filled triangles (black), the
70\,\micron\ bright far-infrared sources are marked with "+" (black)
and the 70\,\micron\ dark FIR sources  are marked with asterisks. The
positional offsets in arcseconds on the two axes are relative to the
center ($\alpha_{2000}$: 14$^{\rm h}$45$^{\rm m}$20.4$^{\rm s}$,
$\delta_{2000}$: -59\arcdeg52\arcmin03\farcs5).\label{fig_overlay}}
\end{figure*}

The velocity-integrated \CII\ map shows three bright emission regions,
one centered on the  bright radio continuum emission in the southern
part of the ridge, one to the southeast of it, and one to the east,
skirting the S111 bubble (Fig.\,\ref{fig_overlay}). All three emission
regions are connected by faint \CII\ emission arising from diffuse PDR
gas.  When compared with the 8\,\micron\ Spitzer image the \CII\
emission is clearly seen to delineate the rims skirting the bubbles
S109 and S111 in particular and matches well with the emission features
in the 8\,\micron\ which are due to FUV-excited PAH molecules in the
PDRs. The \CII\ emission also matches well with the far-infrared dust
continuum emission at 70\,\micron. The tracers of cold dust
(870\,\micron\ ATLASGAL) and ambient molecular material (\thCO(2--1)
and \CeiO(2--1)) primarily show the  molecular ridge. The southeastern
\CII\ peak is detected in both \thCO\ and \CeiO, suggesting a high
column density of gas in the region.  The HCO$^+$(1--0) map tracing
densities in excess of 10$^5$\,\cmcub\ shows that this line is detected
in the three prominently \CII-emitting regions, and the emission
continues to increase to the north toward the denser part of  the
molecular ridge.  The region hosts a large number of far-infrared
sources, some of which are dark even at 70\,\micron,  and there are
quite a few additional cold clumps detected in the 870\,\micron\
ATLASGAL data (marked on Fig.\ref{fig_overlay}).  The detection of
HCO$^+$(1--0) particularly toward the easternmost parts of the northern
rim surrounding S111 and the detection of associated far-infrared
sources are clear indications of ongoing star formation in dense
clumps.

\begin{figure*}
\centering
\leavevmode
\includegraphics[width=0.6\textwidth]{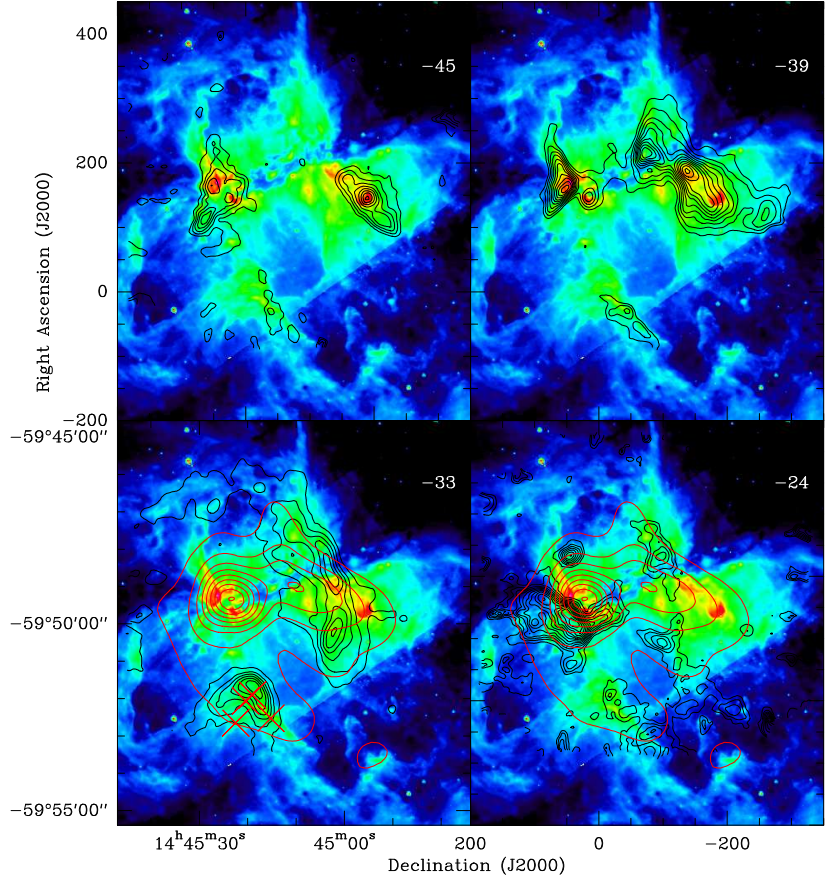}
\caption{Maps of \CII\ emission (contours) in 1\,\kms\ wide channels
centered at velocities indicated in the panels overplotted on the
Spitzer 8\,\micron\ continuum image.  Contours (red) in the  bottom two
panels show the 843\,MHz radio continuum observed by SUMSS and are drawn
from 0.2 to 4.2\,Jy\,beam$^{-1}$ in steps of 0.5\,Jy\,beam$^{-1}$ (for a
beam size of 52\arcsec$\times$45\arcsec).
\label{fig_shellvel}}
\end{figure*}

\section{Velocity distribution of \CII-emitting gas}

\subsection{Velocity-Channel Maps}

Based on velocity-channel maps (Fig.\,\ref{fig_cpluschan}) we identify
four distinct velocity ranges of the \CII\ emission.
Figure\,\ref{fig_shellvel} compares the \CII\ emission in these four
velocity ranges centered at -45, -39, -33 and -24\,\kms\ with the
8\,\micron\ continuum image, which reveals the structure of the region
very well. The spatial segregation of emission in the four velocity
regions indicates the relative motion primarily of the rims of the
bubble S111 and to some extent that of S109. We find that \CII\ shows
multiple velocity components spanning an interval of 20\,\kms\ spread
over about 300\arcsec.  As detailed later in the paper, these velocity
components can be explained by the expansion of the \HII\ region S111.

\subsection{Position--Velocity Diagrams}

\begin{figure}
\centering
\leavevmode
\includegraphics[width=0.5\textwidth]{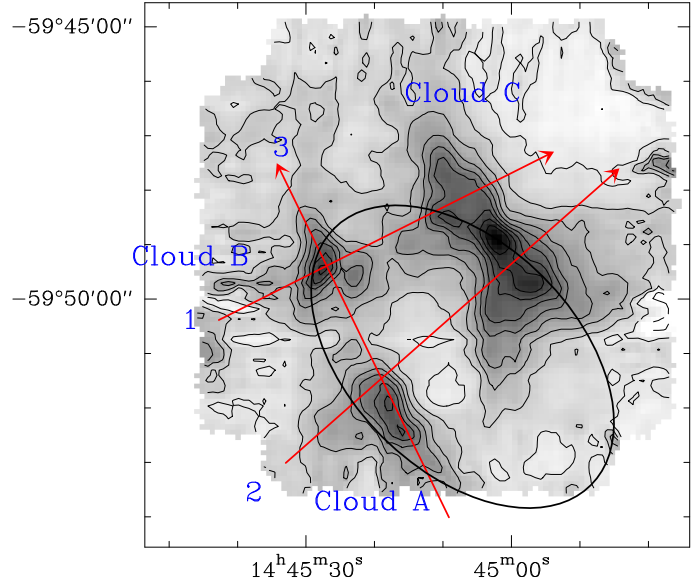}
\caption{Velocity-integrated map of \CII\ emission with intensities
ranging between 10 and 540\,\kms\ (shown by both color and contours). The
contours are separated by 50\,K\,\kms.  Straight lines  (red) with arrows
(marked 1, 2, and 3) show the directions along which position-velocity
diagrams shown in Fig.\,\ref{fig_cpcopv} are derived. Clouds A, B, and C
discussed in the text are also marked. Also shown is an ellipse
approximating the S111 bubble traced by the \CII\ contours. The ellipse
has semi-major and semi-minor axes of 200\arcsec\ and 125\arcsec\
respectively and a position angle of 45\arcdeg\ and is centered at the
position with an offset -100\arcsec,60\arcsec.
\label{fig_cpfil}}
\end{figure}

\begin{figure*}
\plotone{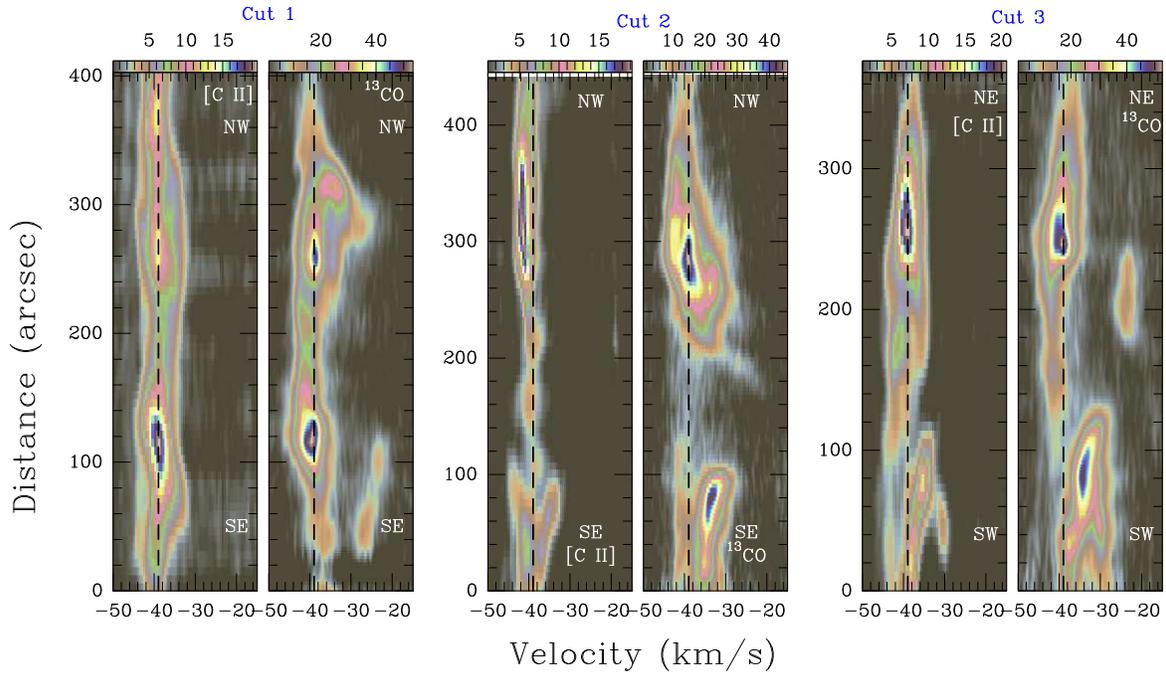}
\caption{Position--velocity maps of \thCO(1--0) and \CII\ emission
along the Cuts 1, 2, and 3 marked in Fig.\ref{fig_cpfil}.
\label{fig_cpcopv}}
\end{figure*}

A more detailed view of the kinematics of the region can be obtained by
studying the position-velocity (p-v) diagrams along a few selected
directions. For ease of reference we identify the \CII\ prominent
emission peaks as clouds A, B, and C (Fig.\,\ref{fig_cpfil}). We
considered directions along the ridge (Cut 1), across the S111 bubble
(Cut 2) and direction perpendicular to the ridge connecting the \CII\
peaks (Cut 3) (Fig.\ref{fig_cpfil}). We compare  the p-v diagrams of
\thCO(2--1) and \CII\ along the three directions. The \thCO(2--1)
emission is representative of the ambient molecular material, while the
\CII\ mostly represents FUV-irradiated neutral gas except for the
immediate neighborhood of the embedded cluster G316.8-00.05
(Fig.\,\ref{fig_cpcopv}). The dotted line in all the p-v diagrams
denotes the centroid velocity of the molecular material, supported also
by the \CeiO(2--1) p-v plots (Fig.\,\ref{fig_cpc18opv}).  Along Cut 1,
in the region near the embedded cluster the \CII\ emission compares
fairly well at velocities around -39\,\kms\ but shows a distinctly
red-shifted component reminiscent of an expanding bubble. The same
feature also shows up as isolated red-shifted blobs on the
northeastern end of Cut 3. We identify this as the interface of the
bubbles S109/S111 that is expanding away from the observer and
propose that the expansion is restricted in the opposite direction by
the high density molecular material that forms the ridge.  Considering
the difference in velocities of the molecular material and the PDR gas
traced by the \CII, we estimate an expansion velocity of around
7\,\kms\ for part of the rim detected around S111. Along Cut 2 across
the bubble S111, the northern \CII\ emission feature is faint and
compact in velocity in \thCO(2--1).  However, \CII\ at the same
position is significantly broadened around the centroid velocity of the
ambient medium, possibly due to the presence of young stellar objects
(YSOs)
that are very bright in the infrared.  Along Cut 2, at the inner edge
of the bubble S111, the \CII\ emission is markedly red-shifted. Along
Cut 3, the southwestern clump shows red-shifted emission, while the
northeastern clump is somewhat blue-shifted relative to the velocity
of the ambient molecular cloud. The velocity distribution seen in the
p-v diagrams of \CII\ is consistent with emission from FUV-irradiated
gas at velocities suggestive of an expanding bubble.

\subsection{Velocity distribution seen in Azimuthally averaged spectra}

\begin{figure}
\centering
\leavevmode
\includegraphics[width=0.35\textwidth]{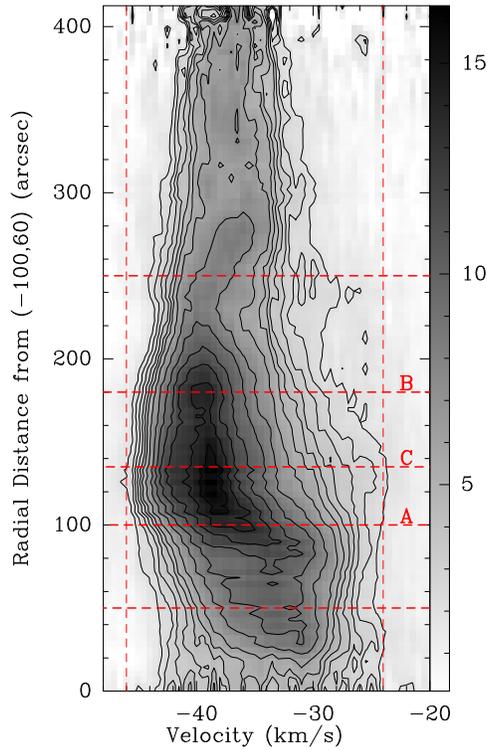}
\caption{Plot of velocity distribution as a function of radial distance from
the position (-100\arcsec,60\arcsec). Each line of the map is the radial profile
for one particular plane of the cube and is created by computing annular 
averages on the data cube. The horizontal dashed lines are drawn at
distances of 50\arcsec, 100\arcsec, 135\arcsec, 180\arcsec, and
250\arcsec\ from (-100\arcsec,-60\arcsec) and show the radial distances of the
clouds A, B and C from the same point.
\label{fig_azav}}
\end{figure}

The analysis of \CII\ emission as a function of velocity in the previous
sections of the paper suggests that \CII\ traces parts of the expanding
bubble S111. The shape of the bubble seen in \CII\ is an ellipse with a
size of 400\arcsec $\times$250\arcsec  centered at $\sim$ -100\arcsec,
+60\arcsec, with a position angle of 45\arcdeg (Fig\,\ref{fig_cpfil}).
In order to improve the signal-to-noise ratio of the high-velocity \CII\
emission, we have computed the \CII\ emission as a function of the
radial distance from the assumed center of the bubble, i.e. at an offset
(-100\arcsec,60\arcsec) (Fig.\,\ref{fig_azav}).  Each line of the map is
the radial profile for one particular plane of the cube and is created
by computing annular averages on the data cube.  This plot shows that
the \CII\ emission is quite broad over the whole nebula, with line
widths in the range of 18 -- 22 \kms{} (Fig.\,\ref{fig_azav}) consistent
with an expanding lobe/bubble with a radius of approximately 220\arcsec.
The \CII\ emission is particularly strong in the interface regions of
clouds A, B, and C. There is a clear velocity gradient between clouds A
and C, with the bulk of the gas being more redshifted in cloud A than in
clouds B and C, likely due to the interaction between the S111 and S109
bubbles. In these radially averaged data the broadest \CII\ lines are
seen in the direction of cloud C, whereas the p-v plots
(Fig.\,\ref{fig_cpcopv}) show that the broadest \CII\ emission is in the
direction of cloud B, which hosts the O-stars responsible for ionizing
the \HII\ region. This can be explained by the fact that the cloud C  is
much more extended in \CII\ than cloud B, so that it dominates the
radial average of the data, which is more sensitive to the extended
emission.

\subsection{Analysis of \CII\ spectra at selected positions}

\begin{figure*}
\plotone{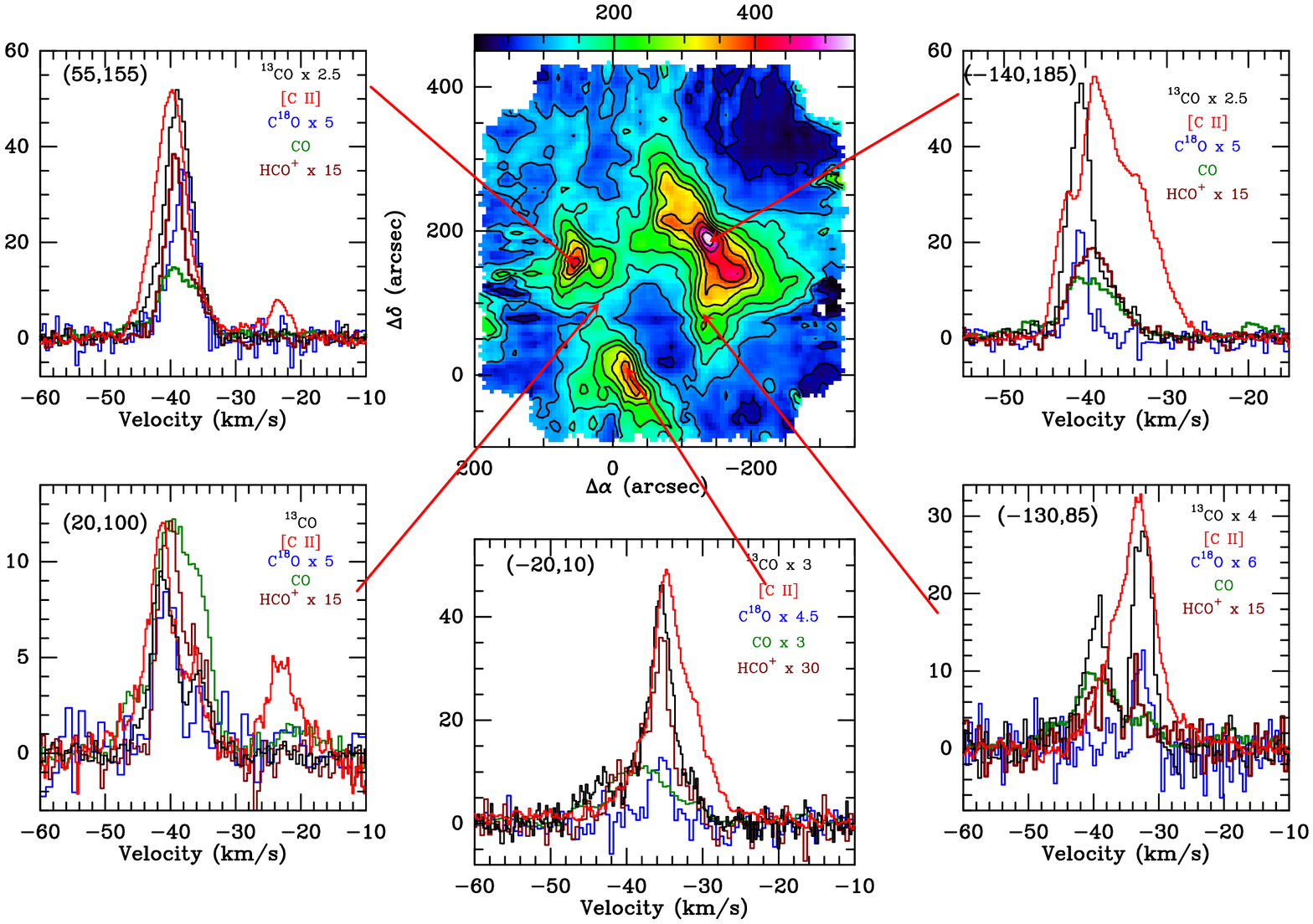}
\caption{Comparison of spectra at selected positions in the observed
\CII\ map (shown in the middle). In each panel the offset of the
position relative to ($\alpha_{2000}$: 14$^{\rm h}$45$^{\rm
m}$20.4$^{\rm s}$, $\delta_{2000}$:
-59\arcdeg52\arcmin03\farcs5) is marked and
\CII\ (black), CO(1--0) (purple), \thCO(2--1) (red), \CeiO(2--1)
(blue), and HCO$^+$(1--0) (brown) spectra are shown. Single- or multiple-component Gaussian fits to
the three spectra at each position were performed and the results are
shown as green continuous curves with details presented in
Table\,\ref{tab_gfit}. The spectra are scaled by the factors mentioned in
the panels for better visibility.
\label{fig_gfit}}
\end{figure*}

Figure\,\ref{fig_gfit} shows a comparison of \CII, \thCO(2--1) and
\CeiO(2--1) spectra at selected positions where \CII\ emission peaks are
observed and which coincide with the locations of Hi-GAL far-infrared
sources. The selected positions also sample the environment in the different
parts of the arc-shaped rim around S111. The \CII\ spectra trace the warm
gas around the embedded FIR sources as well as the gas photodissociated by
the embedded cluster G316.8-0.05. Therefore the emission profiles appear more
complicated than the \thCO\ and \CeiO\ profiles, which are dominated by the
dense molecular gas.   Table\,\ref{tab_gfit} presents the results of single-
or multiple-component Gaussian fits to the \CII, \thCO(2--1) and \CeiO(2--1).
We find that the \CeiO(2--1) profiles are well fitted by a single
velocity component although  the  central velocity appears to vary by more
than 5\,\kms\ across the bubble.  At all positions, one of the components
seen in \CII\ matches with the component seen in \CeiO\ and with one of the two
velocity components typically seen in \thCO. The larger line widths
(5--12\,\kms) seen in \CII\ are likely due to photoevaporative flows of
possibly overlapping PDRs. The broad lines seen in \thCO(2--1) and
HCO$^+$(1--0) likely arise from outflows associated with the young embedded
far-infrared sources.  In particular, the outflow from the embedded YSO is
clearly detected in the \thCO\ and HCO$^+$ spectra of the position with
an offset of (-20\arcsec, 10\arcsec) which is located in cloud A
(Fig.\,\ref{fig_gfit}).

The \CII\ spectra at all positions except toward the center of the star
cluster at (55\arcsec, 155\arcsec) are more redshifted than the
molecular emission.  The hydrogen RRLs observed by \citet{longmore2009}
have central velocities between -42 and -46\,\kms, i.e., they are
significantly blue-shifted compared to the \CII, \thCO, and \CeiO\
spectra.  However, the RRL spectra are much broader than the molecular
lines and \CII.  The line widths are between 20 and 24\,\kms, and
therefore they overlap with some of the \CII\ spectra.  While \CII\ may
originate  from ionized gas as well, in all cases where \CII\ is
detected at  blue-shifted velocities, \thCO(2--1) is also detected,
which implies that the  \CII\ emission is primarily associated with the
neutral gas.

\begin{deluxetable*}{cccccr}
\tablecaption{Results of Gaussian fits to spectra \label{tab_gfit}}
\tablehead{
\colhead{Position} & \colhead{Transition} & \colhead{$\upsilon_{\rm
LSR}$} & \colhead{$\Delta\upsilon_{\rm LSR}$} & \colhead{$I$} & \tkin$^a$\\
\colhead{(arcsec)} & \nocolhead{} & \colhead{(\kms)} &
\colhead{(\kms)} & \colhead{(K\kms)} & (K)
}
\decimalcolnumbers
\startdata
(20, 100) & \CII & -41.52$\pm$0.07 & 5.07$\pm$0.21 & 59.97$\pm1.89$ & 42\\
         &      & -36.04$\pm$0.14 &  2.98$\pm$0.36 & 27.21$\pm1.47$ & \\
         &      & -23.88$\pm$0.15 &  5.55$\pm$0.35 & 12.70$\pm1.50$ & \\
         &\thCO(2--1) & -38.92$\pm$0.49 &6.90$\pm$0.28 & 25.37$\pm1.42$ & \\
         &            & -40.03$\pm$0.18 &6.77$\pm$0.20 & 66.14$\pm1.55$ & \\
         &\CeiO(2--1) & -39.40$\pm$0.24 &6.71$\pm$0.43 & 18.25$\pm1.22$ & \\
(55, 155) & \CII\ & -39.91$\pm$0.25 & 6.85$\pm$0.25 & 281.08$\pm3.84$ & 90\\
         &        & -40.11$\pm$0.25 & 3.30$\pm$0.25 & 47.03$\pm3.84$ &\\
         &        & -22.55$\pm$0.25 & 5.43$\pm$0.25 & 35.76$\pm3.84$ &\\
         &\thCO(2--1) & -38.69$\pm$0.5 &3.64$\pm$0.5 & 39.88$\pm2.93$ &\\
         &            & -39.57$\pm$0.5 &6.83$\pm$0.5 & 73.22$\pm2.93$ &\\
         &\CeiO(2--1) & -38.05$\pm$0.08 &4.40$\pm$0.19 & 31.08$\pm1.08$ &\\
(-20, 10) & \CII\     & -34.91$\pm$0.03 & 2.38$\pm$0.08 & 47.51$\pm2.39$ & 88\\
         &           & -33.94$\pm$0.03 & 8.39$\pm$0.09 & 271.40$\pm2.75$ & \\
         &\thCO(2--1) & -35.41$\pm$0.04 &3.16$\pm$0.12 & 38.82$\pm1.80$ & \\
         &            & -38.04$\pm$0.31 &13.75$\pm$0.60 & 48.01$\pm2.47$ & \\
         &\CeiO(2--1) & -35.55$\pm$0.14 &2.96$\pm$0.35 &  8.98$\pm0.88$ & \\
(-140, 185) & \CII\ & -38.98$\pm$0.02 & 3.63$\pm$0.08 & 144.18$\pm6.36$ & 93\\
         &        & -34.71$\pm$0.09 & 8.02$\pm$0.13 & 297.54$\pm6.69$ &\\
         &        & -42.64$\pm$0.02 & 2.67$\pm$0.05 & 67.73$\pm1.43$ &\\
         &\thCO(2--1) & -36.31$\pm$0.21 &4.55$\pm$0.48 & 17.98$\pm1.83$ &\\
         &            & -40.70$\pm$0.03 &3.22$\pm$0.07 & 69.32$\pm1.71$ &\\
         &\CeiO(2--1) & -40.71$\pm$0.06 &2.11$\pm$0.15 & 10.15$\pm0.59$ &\\
(-130, 85) & \CII\ & -34.09$\pm$0.25 & 4.37$\pm$0.25 &  8.03$\pm2.56$ & 69\\
         &        & -34.03$\pm$0.25 & 9.22$\pm$0.25 & 195.84$\pm2.56$ & \\
         &        & -32.59$\pm$0.25 & 3.34$\pm$0.25 &  41.51$\pm2.56$ & \\
         &\thCO(2--1) & -32.51$\pm$0.06 &3.28$\pm$0.13 & 26.00$\pm0.92$ & \\
         &            & -39.55$\pm$0.13 &4.72$\pm$0.39 & 21.04$\pm1.19$ & \\
         &\CeiO(2--1) & -32.58$\pm$0.18 &2.50$\pm$0.37 & 5.25$\pm0.71$ & \\
\enddata
\tablecomments{$^a$ Planck-corrected peak temperature of the \CII\ spectrum
assumed to be optically thick.}
\end{deluxetable*}

\section{Column Density and Mass of the \CII-emitting gas}

The velocity distribution of the region with the strongly red-shifted
component detected in the \CII\ emission does not allow for the detection of
the much fainter hyperfine component  of \thCII\ line, which lies at
$\upsilon = +11$\,\kms\ relative to the \CII\ line. However, in recent
velocity-resolved observations, the \CII\ emission from most Galactic PDRs
has been found to be optically thick \citep{Mookerjea2021, Guevara2020,
Mookerjea2019}. The PDR gas in the G316.75 region being studied here is
reasonably bright and  high-density clumps are seen in the \CII\ emission as
well.  We thus consider the \CII\ emission to be optically thick and use the
Rayleigh-Jeans corrected peak brightness temperature of \CII\ spectra as an
estimate of the kinetic temperature of the PDR gas (Column 6 of
Table\,\ref{tab_gfit}) and also for the estimation of $N$(\cplus) from the
observed velocity-integrated \CII\ intensity. We first consider the
contribution of ionized gas close to the \HII\ region. Comparison of the
\CII\ emission with the H57$\alpha$ \citep{longmore2009} emission map
suggests that the peak of the \CII\ emission is shifted by about 10\arcsec,
which is within the 14\arcsec\ beam of the \CII\ data.  \citet{Watkins2019}
explained the small ionized gas fraction in the ridge G316.75 as as a
consequence of the large electron ($n_e$=5000\,\cmcub) and H$_2$ gas
densities in the immediate vicinity of the ionizing high-mass stars. If we
assume $n_e$=5000\,\cmcub\ and follow the formulation presented by
\citet[][Equation 7]{Pabst2017} to estimate the contribution of the ionized
gas to the \CII\ intensities close to the \HII\ region we obtain an
integrated \CII\ intensity of 497\,K\,\kms.  The \CII\ intensity estimated to
arise from the ionized gas is therefore comparable to the peak \CII\
intensity of 437\,K\,\kms\ found at this position.  However, unlike the
prediction of the model constructed by \citet{Watkins2019}, the H57$\alpha$
emission suggests that such high-density ionized gas to be confined to within
30\arcsec\ of the peak of ionized emission. At the other extreme ($n_e\sim
60$\,\cmcub{}) of the range of electron densities proposed by
\citeauthor{Watkins2019}, the \CII\ emission would be only 6\,K\,\kms, far
less than the faintest level (180\,K\,\kms) level of \CII\ emission in the
immediate vicinity of the \HII\ region (Fig.\,\ref{fig_cpfil}) . Thus, we
conclude that while for the peak position of the \HII\ region there is a
significant contribution of the ionized gas to the \CII\ emission, the more
extended emission is dominated by the neutral PDR gas.

In the neutral region, where \cplus--H and \cplus--H$_2$ collisions
dominate, we estimate $N$(\cplus) (in \cmsq) from the observed integrated
\CII\ intensities using Eq. (26) from \citet{Goldsmith2012} as
follows:

{\small
\begin{equation}
N(C^+) = {\rm 2.91\times 10^{15}\left[1+0.5{\rm
e^{91.25/T_{kin}}}\left(1+\frac{A_{ul}}{C_{ul}}\right)\right] \int T_{\rm mb}d\upsilon}
\end{equation}
}

where $A_{\rm ul}$ = 2.3$\times 10^{-6}$\,s$^{-1}$, \tkin\ is the gas
kinetic temperature, the collision rate is $C_{\rm ul}$ = $R_{\rm ul}n$
with $R_{\rm ul}$ being the collision rate coefficient with H$_2$ or
H$^0$, which depends on $T_{\rm kin}$, and $n$ is the volume density of
H. At \tkin\ = 100\,K for \cplus--H collisions $R_{\rm ul}$ = 7.6$\times
10^{-10}$\,cm$^3$\,s$^{-1}$ and $R_{ul}$ = 3.8$\times
10^{-10}$\,cm$^3$\,s$^{-1}$ for \cplus--H$_2$ collisions. Since the
critical density of the \CII\ transition is  $n_{\rm cr}$=3000\,\cmcub,
and since it is likely that most of  the  \CII\ detected could be at
such densities along with some emission arising from clumps with
densities exceeding 10$^5$\,\cmcub, we assume a density of
10$^4$\,\cmcub\ to estimate the $N$(\cplus) density distribution of the
region. Assuming a kinetic temperature of 100\,K for \cplus\  and a
density of $n$=10$^4$\,\cmcub, and excitation from  \cplus--H$_2$
collisions, the column density  is found to lie in the range
(0.7--4.5)$\times 10^{18}$\,\cmsq. A combination of slightly different
\tkin\ and density and the inclusion of \cplus--H collisions would
result in an uncertainty of no more than 30\% in the column density.
Based on the values of $A_{\rm ul}$ = 2.3$\times 10^{-6}$\,s$^{-1}$ and
$n_{\rm cr}$ at 100\,K and C/H = 3$\times 10^{-4}$, the \CII\ line is
expected to be optically thick for $N$(\cplus) = 4$\times
10^{17}$\,\cmsq\ \citep[Table 2.7 in][]{TielensBook}.  The estimated
values of $N$(\cplus) thus validate our assumption of the \CII\ line
being optically thick.

We estimate the mass of the \CII-emitting gas, which is dominated by the
three regions named clouds A, B and C (Fig.\ref{fig_cpfil}).  For the
observed \CII\ intensities, a distance of 2.7\,kpc  and a C/H abundance
ratio of 1.3$\times10^{-4}$ of which approximately 50\% of carbon
resides in \cplus, we find the total molecular H$_2$ masses of clouds A,
B and C to be 1000, 640, and 2500\,\msun\ respectively. We have also
estimated the mass of the molecular gas in clouds A, B, and C using the
$N$(H$_2$) maps derived by \citet{Samal2018} based on Herschel
observations. We find the H$_2$ mass estimated from dust continuum
observations in clouds A, B, and C to be 2000, 3400, and 5000\,\msun\
respectively. The estimate of the total molecular gas mass within the
lowest \CII\ contour shown in Fig\,\ref{fig_cpfil} is 6900\,\msun\ from
\CII\ data and 18,100\,\msun\ from the dust continuum. Thus, we see that
in clouds A and C, which are bright in far-infrared dust continuum as
well as \CII, about 50\% of the mass is traced by the \CII\ emission,
whereas in cloud B, in which the ionizing stars reside,  only 18\% of
the mass is traced by \CII.  The lower fraction of molecular mass being
traced by \CII\ in this region is likely due to the high densities in
the ridge, which shields it from the FUV radiation, unlike the rest of
the cloud. The \CII\ emission seen from the ridge is likely due to dense
the ionized gas near the center of the \HII\ region and diffuse PDR
further out.  This indication of higher density material to the north
along the ridge is corroborated by the analysis presented by
\citet{Watkins2019} in support of the lower than expected effect of the
embedded cluster on the ridge itself.  We propose that the bipolar \HII\
region clears off part of the bubble, and the FUV radiation leaking
through the bubble creates PDRs on the surface of the G316.75 ridge.

\section{Estimate of Mechanical feedback from the stars on S111}

\subsection{Energy budget in S111}

The triple bubble system of S109, S110 and S111 has putatively been
created by the massive stellar cluster embedded in the G316.75 ridge.
While the three bubbles are connected through their history of
formation, in the present configuration the \HII\ regions appear to be
segregated by the dense ridges or shells of swept-up gas and hence can
be studied in isolation. Our study focuses on the mechanical and
radiative feedback from the embedded stars.

We first estimate the role of mechanical feedback from the ionizing
stars by comparing the kinetic energy of the shell to the 
the thermal energy of the ionized gas in S111.
The \CII\ observations suggest that the denser parts of the clouds A and
C represent part of the shell created by the expanding \HII\ region.
These  have velocities red-shifted by about 5--6\,\kms\ relative to
cloud B,  which hosts the ionizing stars (Table\,\ref{tab_gfit},
Fig.\ref{fig_cpcopv}). Additionally more diffuse PDR material also
traced by \CII\ are red-shifted by 15\,\kms\ at the boundary between the
bubbles S109 and S111. Thus, the shells of S111 appear to move at
relatively slow speeds, particularly when contrasted with the shells that
have been observed in Orion, RCW\,120, and RCW\,49
\citep{Pabst2020,Luisi2021,Tiwari2021}. The high-density  material
toward the northern part of the ridge seen in the molecular line
observations most likely contributes to the slowing down of the
expansion of the \HII\ region and the shell.  We estimate the kinetic
energy of the clouds A and C with a mass of 7000\,\msun\ moving with a
velocity of 7\,\kms\ relative to  B to be 3.4$\times 10^{48}$\,erg.

The \HII\ region S111 has a radius of $\sim$ 160\arcsec\ (2\,pc at a
distance of 2.7\,kpc) (Fig.\,\ref{fig_cpfil}) and \citet{Watkins2019}
estimated  a temperature $T_{\rm HII}$ = 6600\,K and  density $n_{\rm
e}$=60--5000\,\cmcub. The total thermal energy of the S111 \HII\
region can be written as ${\rm E_{th} = \frac{3}{2} n_e
\frac{4\pi}{3}\,R^3 k_B T_{\rm HII}}$, where ${R}$ is
the radius of the \HII\ region and $k_{\rm B}$ is the Boltzmann
constant. We estimate the $E_{\rm th}$ to range between 8.1$\times
10^{46}$ and 6.7$\times 10^{48}$\,erg for $n_{\rm e}$=60 and 5000\,\cmcub\
respectively.

Thus, depending on the actual electron density of the \HII\ region, the
kinetic energy of the molecular shells primarily detected in \CII\
is between $\sim$ 0.5--40 times the thermal energy of the \HII\ region. 

\subsection{Pressure balance of S111}

The impact of high-mass stars on the ambient molecular cloud can be
further quantified in terms of the pressures associated with direct
radiation ($P_{\rm dir}$), dust-processed radiation ($P_{\rm IR}$),
photoionization heating ($P_{\rm ion}$) and shock heating from stellar
winds ($P_{\rm X}$).  No X-ray mapping observations are available for
the \HII\ region associated with S111, hence $P_{\rm X}$ could not be
estimated. Thus, we next consider the pressures associated with the
other feedback mechanisms from the high-mass stars in the neighborhood
of S111.

The pressure due to the photoionization heating of the gas in the
region, $P_{\rm ionized}/k_{\rm B} = n_{\rm e} T_{\rm HII}$ is estimated
to be between (0.4--33)$\times 10^6$\,K\,\cmcub\  for electron densities
of 60--5000\,\cmcub\ respectively.

Following \citet{Olivier2021}, we define the direct radiation
pressure ($P_{\rm dir}$) as the momentum available to drive motion in
the shells in the \HII\ region at a radius $R$ from the central stars
(Eq. 3)

\begin{equation}
{\rm P_{dir}} = {\rm \frac{3L_{bol}}{4\pi R^2c}},
\end{equation}

where $L_{\rm bol}$ is the bolometric luminosity of central stars.  The
total bolometric luminosity of the two central ionizing stars, one O7V
and the other O6V, is 3.5$\times 10^5$\,\lsun\ and the  total production rate of
ionizing photons is 1.56$\times 10^{49}$\,s$^{-1}$.  The direct
pressure ($P_{\rm dir}/k_{\rm B}$) at the location of the inner edge of
the shell located at a radius of 160\arcsec\ (2\,pc) from the stellar
cluster is therefore  9.3$\times 10^5$\,K\,\cmcub.

In another feedback mode, stellar radiation is absorbed by dust and
thermally reradiated in the IR wavelengths, and this induces a
dust-processed radiation pressure that is given by {\bf \citep[Eq. 2
in][]{Olivier2021}}

\begin{equation}
{\rm P_{IR}} = \frac{1}{\rm 3V}a{\rm T_d^4}\frac{4\pi}{3}{\rm
(R_{\rm shell}^3 -R^3)},
\end{equation}

where $R_{\rm shell}$ is the outer radius of the shell, $R$ is the
inner radius of the shell, which is the same as the radius of the \HII\
region, and $V$ is the volume of the \HII\ region. Based on the dust
temperatures $T_{\rm d}$   derived by \citet{Samal2018}, we assume the
dust temperature in the shell to be around 24\,K. From the
position-velocity plots (Fig\,\ref{fig_cpcopv}) we estimate the
thickness of the shell to be 20\arcsec\ (0.3\,pc), resulting in a
dust-processed radiation pressure $P_{\rm IR}/k_{\rm B}$ of  6$\times
10^6$\,K\,\kms.

Finally, we estimate the thermal pressure ($P_{\rm PDR}$) of
\CII-emitting PDR gas. We assume the hydrogen volume density of the PDR
to be $n_{\rm H}$= 3$\times 10^3$\,\cmcub, which is the critical density
($n_{\rm cr}$) for the \CII\ transition, and the gas temperature ($T_{\rm
PDR}$) to be 100\,K based on our estimates in Section\,6. Thus ${\rm
P_{\rm PDR}} = n_{\rm H} k_{\rm B} T_{\rm PDR}= 3\times 10^5$\,K \kms.  It
is likely that the value of $P_{\rm PDR}$ is an underestimate, since we
have considered only the diffuse component of the PDR and it could
easily be a factor of three  larger, in which case it would be
comparable to $P_{\rm dir}$.

We find that among all factors contributing to pressure, $P_{\rm ion}$
is the most important, and depending on the electron density it is of
similar in magnitude to $P_{\rm IR}$ of the shell around the \HII\
region. In this analysis, we have not considered the dust-processed
radiation pressure inside the \HII\ region since there is very faint or
almost no dust continuum detected from within the \HII\ region even in
the SPIRE 250\,\micron\ maps. We find that $P_{\rm dir}$ and $P_{\rm
PDR}$ are lower than $P_{\rm IR}$ by factors of 6 and 20, respectively.
\citet{Olivier2021} considered resolved compact \HII\ regions and found
that most of the sources are dominated by $P_{\rm IR}$, and the median
$P_{\rm dir}$ and $P_{\rm ion}$ are smaller than the median $P_{\rm IR}$
by factors of 6 and 9 respectively. Based on the radial dependence of
pressure terms, these authors concluded that the \HII\ regions
transition from $P_{\rm IR}$-dominated to $P_{\rm ion}$-dominated at
radii of $\sim 3$\,pc.  Our conclusions for the S111 \HII\ region,
which has a radius of about 2\,pc with a shell of about 0.3\,pc
thickness, are consistent with the results obtained by
\citet{Olivier2021}.

\section{Discussion}

The velocity-resolved observations of \CII\ have enabled a tomographic
study of the expanding shell associated with the bubble S111 formed by
the embedded cluster G316.8-00.05. The expansion of the bubble at $\sim
7$\,\kms\ is primarily observed in the red-shifted velocities with no
blue-shifted counterparts. This suggests that the bubble is able to
expand more freely in a direction away from the observer, possibly due to
the presence of higher density material between the rim of S111 and
the observer.  The \HII\ region associated with S111 plays a key role
in providing the pressure for the expansion of the shell in the ambient
molecular medium. We estimate ionization of hydrogen ($P_{\rm ion}$) and
dust-reprocessed radiation ($P_{\rm IR}$) to be the primary contributors
to the pressures in the region. We find that the kinetic energy of
part of the molecular shell associated with clouds A and C is 
0.5--40 times the total thermal energy of the \HII\ region.  We do not
see any evidence of injection of energy by stellar winds from the
massive stars; the moderate expansion velocity of the shell can be well
explained by the smooth expansion of the \HII\ region.  This is in
contrast to several recent studies of the feedback of massive stars on
their environment, which imply a significant role of the stellar winds in
the Orion veil shell, RCW\,120, and RCW\,49
\citep{Pabst2020,Luisi2021,Tiwari2021}.  There is a distinct difference
between the ionizing stars in the G316.75 region and regions studied by
the above-mentioned works. The ionizing stars in Orion, RCW\,120, and
RCW\,49 have already cleared the surrounding region, whereas the
G316.80-0.05 cluster is still deeply embedded in dense molecular
material with a line-of-sight visual extinction exceeding 100\,mag. This
suggests that the impact of the stellar winds, if any, from G316.8-0.05
could have also been contained within a very small region due to the
high densities of the region.

The dense rim created by the expansion of the bubble has been
studied in terms of the three clouds A, B, and C, with cloud B being
coincident with the ridge G316.75, which is dark in the mid-infrared
and hosts the massive stellar cluster that gave rise to the triple
bubble system.  Clouds A and C are created by the compression
of molecular material by the expanding bubble and contain embedded
far-infrared sources that are indicative of formation of a new
generation of stars.  The two \CII\ emission peaks in cloud B are
associated with far-infrared sources detected at $\lambda >
70$\,\micron\ as well as with 870\,\micron\ ATLASGAL sources.  The \CII\
peak in the cloud C coincides with the ATLASGAL source
G316.768-00.026, which remains undetected at $\lambda
<160$\,\micron, consistent with a YSO at a very early stage of
evolution.  Most of the YSOs identified based on mid-infrared
colors \citep{Samal2018} lie along the northern part of the region,
with one lying very close to the head of cloud A, one close to
cloud B, and two in cloud C. One of the reasons for the
nondetection of YSOs in the region is attributed to the high
degree of obscuration along this line of sight. 

Cloud A in particular appears to have a structure similar to the
pillars created by the UV erosion of molecular material
(Fig.\,\ref{fig_shellvel}).  At the head of cloud A
(-20\arcsec,10\arcsec) an outflow is detected in the form of a
pronounced blue wing in the \thCO(2--1) and HCO$^+$(1--0) spectra
(Fig.\,\ref{fig_gfit}) arising from the embedded YSO. The outflow
coincides with the ATLASGAL source G316.778-00.096, which is also
detected in the far-infrared Hi-GAL images with its intensities peaking
at 160\,\micron.  \citet{elia2017} have identified this source as
HIGALBM\,316.7799-0.0942 and have derived a mass of 234.6\,\msun, a
bolometric luminosity of 1187\,\lsun, and a bolometric temperature of
38.6\,K. While no other signs of star formation have yet been reported,
the detection of broad spectra consistent with an outflow, high
densities (based on detection of HCO$^+$(1--0))  along with the
presence of a compact far-infrared source are clearly indicative of
ongoing star formation activity.  Based on the mass estimated for the
continuum source and the distance to the region, it is possible that a
star cluster is being formed at the head of cloud A.  The \HII\ region
of S111 is estimated to have an age of 2\,Myr, which when compared
with typical formation timescales of high-mass stars implies that the
star formation activity in cloud A is more recent and hence likely
triggered by the expansion of S111.

\section{Conclusions}

We have used velocity-resolved observations of the
$^2$P$_{3/2}\rightarrow ^2$P$_{1/2}$ transition of \CII\ and the
$J$=2--1 transition of \thCO\ and \CeiO\ to study the infrared bubble
S111, which lies in the region surrounding the southern part of the
infrared dark ridge G316.75.  The \CII\ emission from the region
primarily arises from the shells irradiated by the embedded cluster
G316.80--0.05, part of this photodissociated gas in the shell in cloud C
not being detected in \thCO(2--1) and \CeiO(2--1). In the dense ridge
G316.75 the \CII\ emission arises due to photodissociation of the
molecular material by the stellar radiation leaking out into the bubble
and not through the dense ridge. The \CII\ emission traces the
low-density PDR gas and shows dense clumps that emit HCO$^+$(1--0) as
well.  However, in the absence of any other PDR tracer it is not
possible to further constrain the properties of the PDR gas.

The velocity information of the \CII\ data in particular enabled us to
decipher the morphology of the expanding shell and evaluate the feedback
from the massive star cluster on the surrounding ISM. The expanding
shell is primarily propelled by the \HII\ region created by the embedded
cluster G316.8-0.05.  Azimuthally averaged \CII\ data were used to
conclusively show that the entire velocity structure seen in \CII\ can
be explained by layers of a single shell expanding with a moderate
velocity of $\sim 7$\,\kms\ around S111. We conclude that the
kinetic energy of the expanding part of the shell is comparable to or
larger than the thermal energy of the \HII\ region. We have also
quantified the contribution of the stellar radiation to the pressure in
the region and conclude that the thermal pressure of the \HII\ region
and the dust-processed radiation pressure are the most significant
components.  We present here a clear evidence of star formation
triggered by the compression of the ISM due to the expanding bubble
S111.

\begin{acknowledgments}

The author acknowledges the insightful discussions with Dr G.  Sandell
during the writing of the manuscript.  This work was supported by
funding by the Department of Atomic Energy, Government of India, under
Project Identification No. RTI 4002.  Based on observations made with
the NASA/DLR Stratospheric Observatory for Infrared Astronomy (SOFIA).
SOFIA is jointly operated by the Universities Space Research
Association, Inc.  (USRA), under NASA contract NAS2-97001, and the
Deutsches SOFIA Institut (DSI) under DLR contract 50 OK 0901 to the
University of Stuttgart. The development of GREAT was financed by the
participating institutes, by the Federal Ministry of Economics and
Technology via the German Space Agency (DLR) under Grants 50 OK 1102,
50 OK 1103 and 50 OK 1104 and within the Collaborative Research Centre
956, sub-projects D2 and D3, funded by the Deutsche
Forschungsgemeinschaft (DFG). This research has made use of the VizieR
catalog access tool, CDS, Strasbourg, France.  The original description
of the VizieR service was published in A\&AS 143, 23.  The ATLASGAL
project is a collaboration between the Max-Planck-Gesellschaft, the
European Southern Observatory (ESO) and the Universidad de Chile. It
includes projects E-181.C-0885, E-078.F-9040(A), M-079.C-9501(A),
M-081.C-9501(A) plus Chilean data.  This paper made use of information
from the SEDIGISM survey database located at
https://sedigism.mpifr-bonn.mpg.de/index.html, which was constructed by
James Urquhart and hosted by the Max Planck Institute for Radio
Astronomy. 

\end{acknowledgments}

%

\bibliography{msg316}{}
\bibliographystyle{aasjournal}

\appendix
\renewcommand\thefigure{\thesection.\arabic{figure}}   
\section{Additional Figures}

Figure\,A.1 presents the channel maps of \CII\ emission discussed in 
Section 5.1. Figure\,A.2  shows the position-velocity diagrams for
\CeiO(2--1) which are discussed in Section 5.2.

\setcounter{figure}{0}
\begin{figure}
\plotone{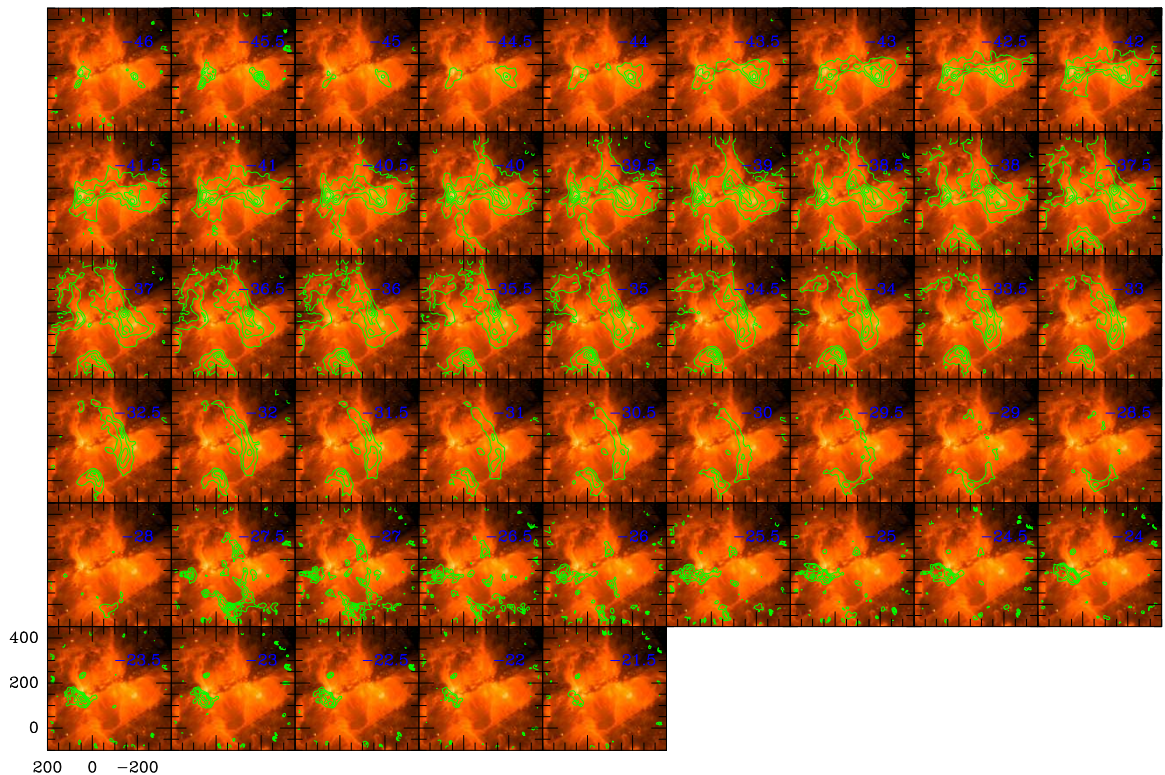}
\caption{Channel maps of \CII\ emission (contours) overlaid on Spitzer
8\,\micron\ continuum image (color).  The positional offsets are relative to
the center ($\alpha_{2000}$: 14$^h$45$^m$20.4$^s$, $\delta_{2000}$:
-59\arcdeg52\arcmin03\farcs5).
\label{fig_cpluschan}}
\end{figure}

\begin{figure}
\plotone{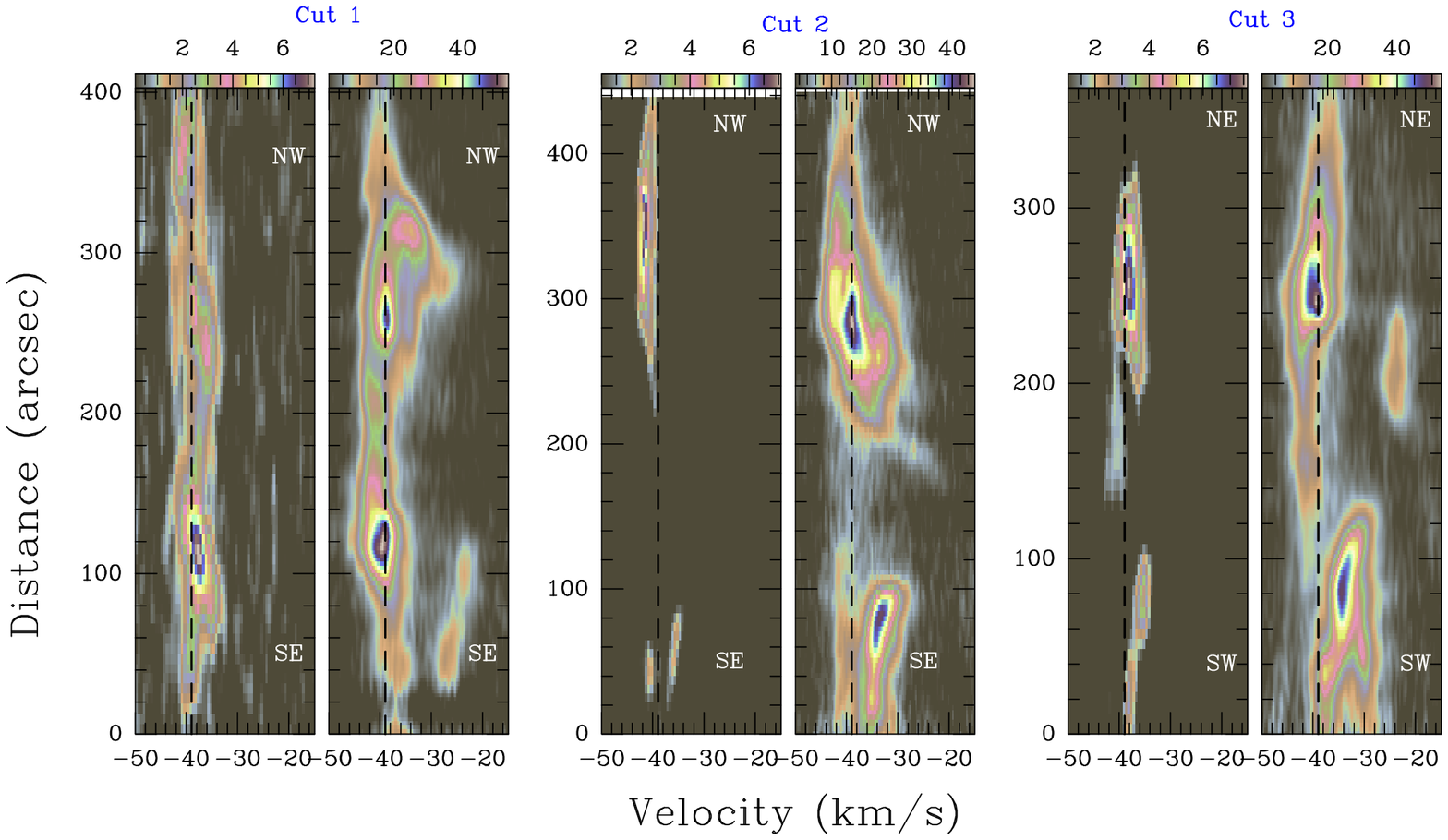}
\caption{Position-velocity maps of \CeiO(1--0) and \CII\ emission
along the Cuts 1, 2 and 3 marked in Fig.\ref{fig_cpfil}.
\label{fig_cpc18opv}}
\end{figure}

\end{document}